\documentclass[conference]{IEEEtran}
\IEEEoverridecommandlockouts
\usepackage{cite}
\usepackage{amsmath,amssymb,amsfonts}
\usepackage{algorithmic}
\usepackage{graphicx}
\usepackage{textcomp}
\usepackage{xcolor}
\usepackage{hyperref}
\def\BibTeX{{\rm B\kern-.05em{\sc i\kern-.025em b}\kern-.08em
    T\kern-.1667em\lower.7ex\hbox{E}\kern-.125emX}}
\begin{document}

\title{COVID-19 Diagnosis from Cough Acoustics using ConvNets and Data Augmentation
}

\author{\IEEEauthorblockN{Saranga Kingkor Mahanta\IEEEauthorrefmark{1}, Darsh Kaushik\IEEEauthorrefmark{2}, Shubham Jain\IEEEauthorrefmark{3}, Hoang Van Truong\IEEEauthorrefmark{4} and Koushik Guha\IEEEauthorrefmark{1}}
\\
\IEEEauthorblockA{\IEEEauthorrefmark{1}Electronics and Communication Engineering Department,
National Institute of Technology, Silchar, India 788010\\
Email: saranga\_ug@ece.nits.ac.in, koushik@ece.nits.ac.in}
\\
\IEEEauthorblockA{\IEEEauthorrefmark{2}Computer Science and Engineering Department, National Institute of Technology,Silchar, India 788010\\
Email: darsh\_ug@cse.nits.ac.in}
\\
\IEEEauthorblockA{\IEEEauthorrefmark{3}Software Engineering Department, Paytm, Noida, India 110096\\
Email: shubhamshrishrimal@gmail.com}
\\
\IEEEauthorblockA{\IEEEauthorrefmark{4}Mathematics in Computer Science, University of Science of HCMC, Ho Chi Minh City, Vietnam 700000\\
Email: hoangvantruong1369@gmail.com}}


\maketitle

\begin{abstract}
With the periodic rise and fall of COVID-19 and countries being inflicted by its waves, an efficient, economic, and effortless diagnosis procedure for the virus has been the utmost need of the hour. COVID-19 positive individuals may even be asymptomatic making the diagnosis difficult, but amongst the infected subjects, the asymptomatic ones need not be entirely free of symptoms caused by the virus. They might not show any observable symptoms like the symptomatic subjects, but they may differ from uninfected ones in the way they cough. These differences in the coughing sounds are minute and indiscernible to the human ear, however, these can be captured using machine learning-based statistical models. In this paper, we present a deep learning approach to analyze the acoustic dataset provided in Track 1 of the DiCOVA 2021 Challenge containing cough sound recordings belonging to both COVID-19 positive and negative examples. To perform the classification on the sound recordings as belonging to a COVID-19 positive or negative examples, we propose a ConvNet model. Our model achieved an AUC score percentage of 72.23 on the blind test set provided by the same for an unbiased evaluation of the models. The ConvNet model incorporated with Data Augmentation further increased the AUC-ROC percentage from 72.23 to 87.07. It also outperformed the DiCOVA 2021 Challenge's baseline model by 23\% thus, claiming the top position on the DiCOVA 2021 Challenge leaderboard. This paper proposes the use of Mel frequency cepstral coefficients as the feature input for the proposed model.
\end{abstract}


\section{Introduction}
The novel Coronavirus or COVID-19  has disrupted life across most countries. The losses are compounding on a daily basis with a sporadic rise in cases. Quite recently, new and stronger variants of the virus are being reported from across the globe, thus making herd-immunity a thing of the distant future. Different vaccination procedures having their intricacies are being carried out across various countries, however, no absolute cure has been discovered yet.

The current testing methods include residual transcription-polymerase chain reaction (RT-PCR) and rapid antigen test (RAT). RT-PCR test involves detection of SARS-CoV-2 like nucleic acid and proteins in the nasal or throat of a person through the use of chemical reagents \cite{covidtest2020}. Although this is an accurate method, the setup for testing is both time and cost-consuming and hence, not scalable given the large number of cases emerging every day. RAT is a quick way of testing, but its reliability is questionable due to the high false negatives. Thus, there is an urgent requirement for economic, efficient, reliable, and methods of early diagnosis that avoid physical contact between the examiner and the patients. 

One of the most common and quickest methods today is temperature check and contact tracing which relies on the fact that the predominant symptom of the COVID-19 is high-temperature fever, in about 89\% of cases \cite{whoreport2020}. However, since the temperature measurement of the human body is dependent on various external factors, the false positive rate is very high. 

 About 67\% of positive COVID-19 cases have shown symptoms of dry cough along with phlegm production in the remaining ones \cite{whoreport2020}. Features of coughs and other vocal sound signals hold information regarding pulmonary health that can be used for COVID-19 infection diagnosis. Detection of discernible symptoms can be performed using machine learning networks trained on recorded cough sounds from healthy and infected subjects. The DiCOVA 2021 Challenge\footnote{\url{https://dicova2021.github.io/}}\cite{muguli2021dicova} was organized to encourage research efforts by creating and releasing an acoustic signal dataset comprising of cough sounds pertaining to both COVID-positive and negative subjects and encouraging researchers to develop detection models and report performance on a blind test set \cite{Muguli}. This paper is the result of our participation in the challenge where We used a Convolutional Neural Network (CNN) architecture \cite{lecun2010convolutional} trained on the provided dataset along with data augmentation that gives an area-under-the-curve (AUC) score percentage of 87.07 on the blind test set.


\section{Related Works}
\label{sec:related_works}

Researchers around the world have shown that it is possible to detect COVID-19 through cough recordings from mobile phones using artificial intelligence (AI) with an accuracy of over 90\% \cite{mit2020}.

Substantial experiments on cough data have been carried out since the COVID-19 outbreak. In \cite{9256562} six features were extracted from data (namely Spectral Centroid, Spectral Rolloff, Zero-Crossing Rate, MFCC, $\Delta$ MFCC, and $\Delta^{2}$ MFCC) and fed to a Long Short-Term Memory (LSTM) network to obtain an accuracy of 97\% for cough sounds, but more importantly have shown that MFCCs were the most important feature based on system accuracy for cough sounds. Another CNN-based approach was followed by Bagad et.al. in \cite{bagad2020cough} for cough samples where a ResNet18 architecture (pre-trained on ImageNet) with short-term magnitude spectrogram as input was able to attain an AUC of 72\%. 

App-based methods for both collections of cough audio samples and detection of COVID-19 were followed in several works. For instance, in \cite{Imran_2020} an Artificial Intelligence-powered smartphone app is developed to provide a preliminary screening for COVID-19 using cough samples.
Using a ConvNet, with Mel-spectrogram images as input reported a sensitivity of 94\% and specificity of 91\%. Crowd-sourced data of cough and breathing sounds were used in \cite{10.1145/3394486.3412865}, from which various frame and segment level features were extracted and combined with features extracted from a pre-trained VGGish network, to be then passed through a logistic regression classifier. The results show an AUC of 80\%. 

Another study used an ensemble of 3 different networks for MFCCs, Mel-spectrograms, and extra features (binary label for the presence of respiratory diseases and fever/myalgia) respectively to predict coronavirus infection with an AUC of 77.1\% \cite{chaudhari2021virufy}. The model also generalized to Latin American and South Asian audio samples. A relatively recent work \cite{fakhry2021virufy} achieved a sensitivity and specificity of 85\% and 99.2\% respectively, along with an AUC of 0.99 for detecting COVID-19 positive cough samples using a ResNet-50 based multi-branch ensemble architecture trained on MFCCs, Mel-frequency spectrograms, and clinical features.

\section{Proposed Method}
\label{sec:proposed}
The entire methodology of our work and the relevant theory is described in detail below.

\subsection{Feature Extraction}
\begin{figure}[!t]
	\centerline{\includegraphics[width=\linewidth]{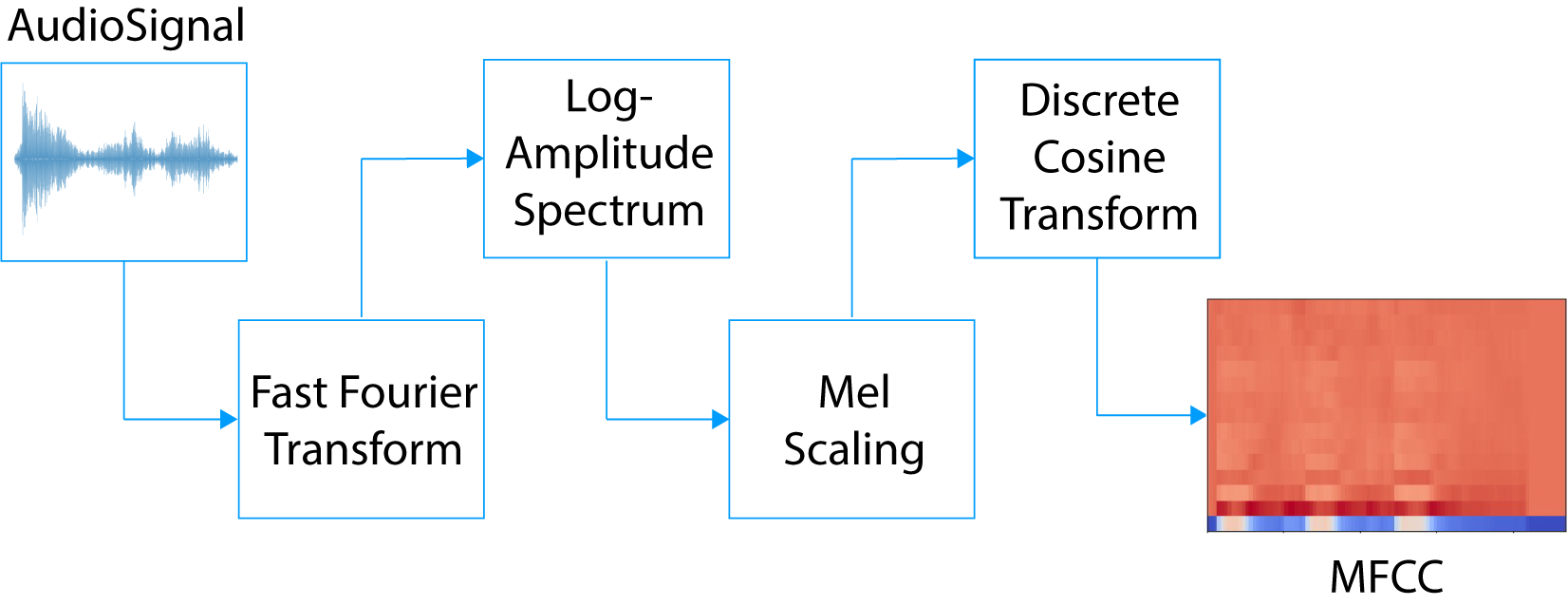}}
    \caption{Extracting Cepstral Coefficients from an audio signal}
    \label{fig:MFCC}
\end{figure}

The information of the rate of change in spectral bands of a signal is given by its cepstrum. It conveys the different values that construct the formants and timbre of a sound. The cepstral coefficients can be extracted using Equation \ref{cepstrum}.

\begin{equation}
   \label{cepstrum}
   C(x(t))=F^{- 1}(log(F[x(t)]))
\end{equation}

Peaks are observed at periodic elements of the original signal while computing the log of the magnitude of the Fourier transform of the audio signal followed by taking its spectrum by a cosine transformation as shown in Fig \ref{fig:MFCC}. The resulting spectrum lies in the quefrency domain \cite{oppenheim2004frequency}. Humans perceive amplitude logarithmically, hence a conversion to the Log-Amplitude Spectrum is perceptually relevant. Mel scaling is performed on it by using Equation \ref{mel} on frequencies measured in Hz.

\begin{equation}\label{mel}
Mel(f)=2595 * log(1+f/700)
\end{equation}

Chaotic dynamics have indicated that nonlinear phenomena exist in vocal signals. Mel frequency cepstral coefficients (MFCCs) \cite{ittichaichareon2012speech} are known to be able to extract critical information from the non-linear and low-frequency regions of sound. Mel-spectrograms can capture small changes in the lower frequency regions since the Mel scale contains unequal spacing in the frequency bands, unlike a typical frequency spectrogram having equally spaced frequency bands \cite{stevens1937scale}. Cough sounds contain more energy in the lower frequencies \cite{knocikova2008wavelet}, consequently MFCCs are an apt representation for the cough sound recordings. It is also widely used in audio detection tasks.

\begin{figure}[!t]
	\centerline{\includegraphics[width=7.4cm]{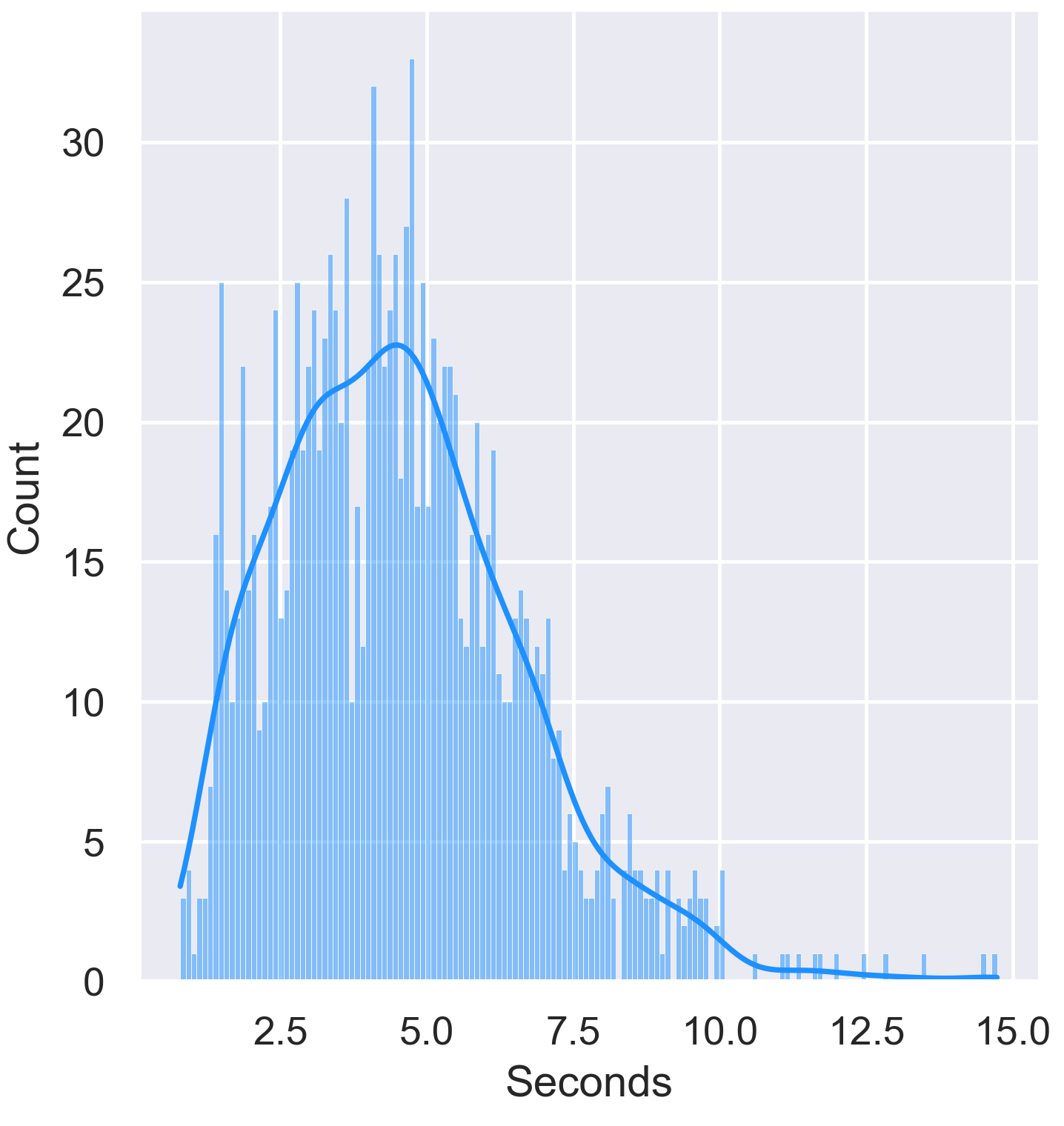}}
    \caption{Distribution of duration of audio samples in the dataset}
    \label{fig:Durations}
\end{figure}

The cough sound recordings of the dataset have audio samples with duration ranging from approximately 0.79 seconds to 14.73 seconds, as shown in the histogram plot in Fig.~\ref{fig:Durations}. Additionally, the input dimensions must be constant across all the training examples to be able to feed into any neural network, thus implying that the MFCC matrices of all the examples need to have a fixed dimensional size. To achieve this, each of the examples must compulsorily have a constant duration resulting in a fixed number of samples when sampled with a constant sampling rate. Out of the 1280 recordings, it was observed that 804, 996, 1130 recordings have duration lesser than 5,6 and 7 seconds respectively. Choosing the right duration for all the recordings is crucial. Choosing a small duration will trim out important information from the sound. On the contrary, choosing the maximum duration i.e 14.73 seconds will result in a tremendous amount of sparse values in the inputs. Considering this trade-off, we chose 154350 samples for each example, which is equivalent to 7 seconds when sampled at 22050 samples/second, as the constant number of samples for all the examples because a credible majority of the recordings have a duration lesser than 7 seconds. Losing valuable information is a graver concern than having too many sparse values since the dataset is relatively small. Moreover, it can be observed from Figure Fig.~\ref{fig:Durations} that only a few recordings have a duration above 7 seconds. Only 150 examples had to be trimmed down while the others had to be padded with zeros to make all 1280 recordings have a constant number of samples of 154350.

After the required trimming and padding of the examples, 15 MFCC coefficients were chosen per frame of each example since the lower end of the frequency axis of the cepstrum contains the most relevant information to our particular task viz. formants, spectral envelope, etc. Moreover, choosing a higher number of cepstral coefficients would proportionally increase the complexity of the model. This may result in a higher variance problem since the dataset is not very large. 15 MFCC coefficients were extracted, using the Librosa \cite{mcfee2015librosa} python library, per frame of each recording. A single frame contained 2048 samples. A hop length of 512 frames was used for the framing window. A duration of 7 seconds was chosen for each cough sound recording. The 7 second sound samples resulted in MFCC matrices having dimensions of 15x302. MFCC features of the sound recordings as its input and produces the probability score of the classification as output.

\subsection{Data Augmentation}

Initially, a simple neural network was trained using the provided dataset as described in Section \ref{sec:setting}. However, due to high class imbalance, present in the ratio of approximately 1:12 with respect to COVID-19 positive-to-negative sounds, the performance was expectedly poor. To overcome this, upsampling  of the COVID-19 positive sound recordings was performed with data augmentation techniques using the Python Audiomentations\footnote{\url{https://pypi.org/project/audiomentations/0.6.0/}}
package. The classes were used with different values of probability parameters to produce a wide range of audio recordings varied with respect to various aspects of the sound signal according to the probability value. And to address them a range of different augmentation functions were used. The following functions were exploited to augment the data:

\begin{itemize}[\IEEEsetlabelwidth{Z}]
\item TimeStretch- The signal is time-stretched without changing the pitch.
\item PitchShift- The pitch of the sound is varied without changing the tempo.
\item Shift- The samples are shifted forwards or backwards, with or without rollover.
\item Trim- The leading and trailing silence of a sound signal is removed.
\item Gain-  The audio is multiplied by a random amplitude factor to increase or reduce the volume.
\end{itemize}

The augmentation functions were selected and used with full consideration of the changes that they might bring to the original sound signal. While augmenting the data, care was taken that the essential features of a sound signal that is the major contributing factor to be diagnosed as COVID-19 positive remain more or less the same, otherwise this upsampling would have rather decreased the performance of the model. Pitch shifts aim to look at how a COVID-19 cough may differ among individuals since pitch is a major component that defines an individual's voice. Trimming the silent parts of the sound helps reduce sparse and needless components. Varying the volume would not affect the contributing features that lead to the diagnosis. Thus, the applied transformations result in physically realizable coughs.

\begin{figure}[!t]
    \centerline{\includegraphics[width=\linewidth]{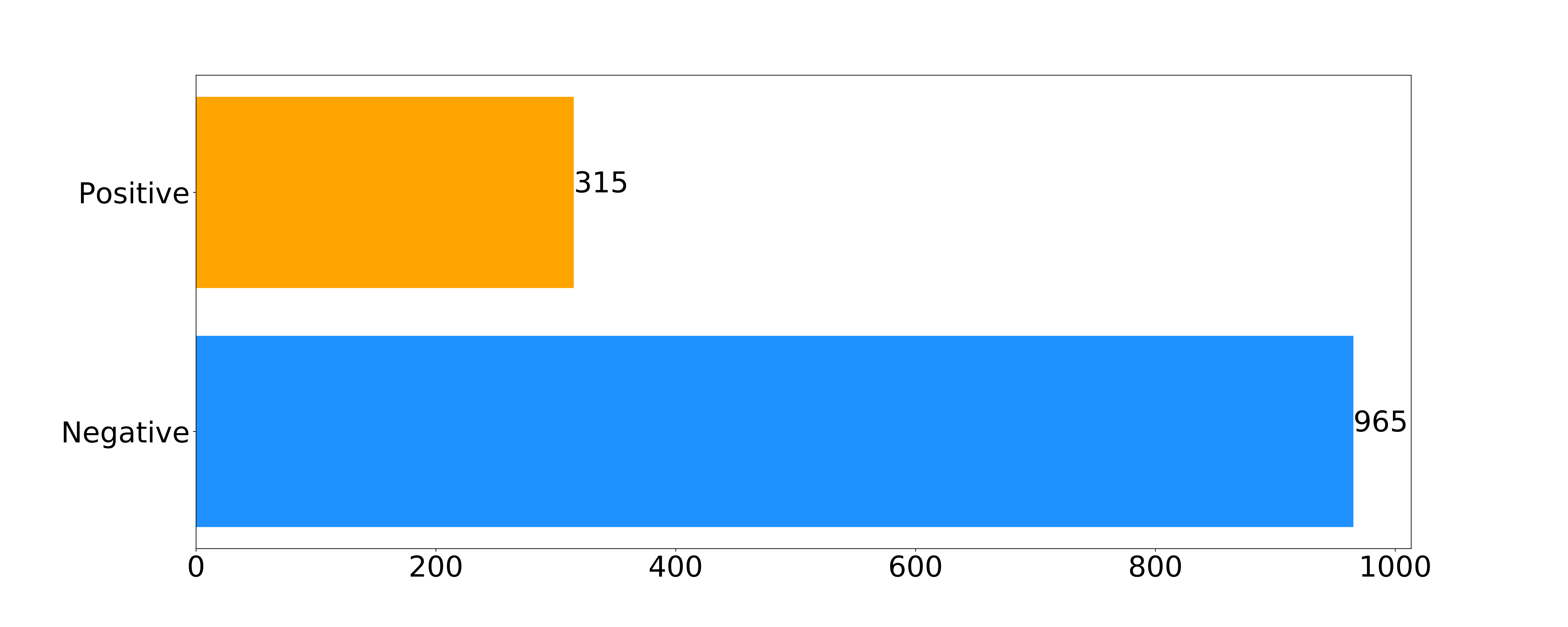}}
    \caption{Class distribution of the augmented dataset}
    \label{fig:aug_distribution}
\end{figure}

The data that was finally used for training after the augmentation had 1280 recordings of cough sound: 965 COVID-19 negative and 315 COVID-19 positive subjects. The class ratio was improved to approximately 1:3, as shown in Fig.~\ref{fig:aug_distribution}.

\begin{figure*}[ht]
	\centering
    \noindent\includegraphics[width=\textwidth, height=4.5 cm]{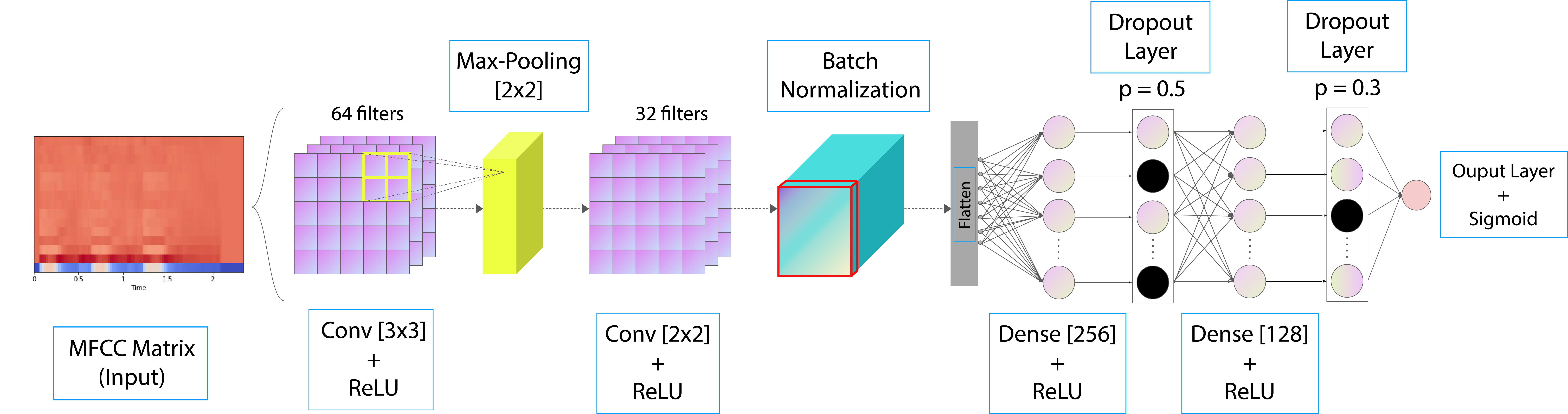}
    \caption{Proposed CNN architecture}
    \label{fig:CNN}
\end{figure*}

\subsection{Model Architecture}
\label{subsec:model}

The sound recordings converted to MFCC matrices were used as input features separately into a multilayer perceptron (MLP) network \cite{wang2017origin}, an LSTM Recurrent Neural Network \cite{sak2014long}, and our CNN. Those literature architectures are shown in Section \ref{sec:related_works}. The evaluation results obtained from the described architectures, including our proposed model and the baseline model, are shown in Table \ref{table:auc}. The baseline model of the DiCOVA challenge used a Random Forest classifier trained with fifty trees. The participants were instructed to submit the output probabilities corresponding to each sound recording name.

The blind test set was provided in the DiCOVA 2021 challenge, along with the data. It consisted of 233 new cough sound recordings on which the models were to perform a classification between the COVID-19 positive and negative examples. The MFCC feature matrices are visually comprehensible and the recent results of CNN image representations show considerable performance gains for the visual recognition tasks. Therefore we propose ConvNets architecture as illustrated in Fig.~\ref{fig:CNN} and elucidated below:

    \begin{enumerate}[\IEEEsetlabelwidth{12)}]
        \item Convolution layer with 64 filters, kernel size of 3x3, stride of 1x1, valid padding followed by ReLU activation function, accepting an input shape of 302x15x1.
        \item Max pooling layer with a pool size of 2x2.
        \item Another Convolution layer with 32 filters, kernel size of 2x2, stride of 1x1, valid padding followed by ReLU activation function.
        \item Batch normalization layer 
        \item The resultant shape was then flattened for the subsequent fully connected layers.
        \item Fully connected layer having 256 units with kernel, bias, and activity regularizers, followed by ReLU activation function.
        \item A dropout layer with a rate of  0.5.
        \item Another fully connected layer having  128 units with kernel, bias, and activity regularizers, followed by ReLU.
        \item Another dropout layer with a rate of 0.3.
        \item Output layer having 1 neuron with Sigmoid activation function.
    \end{enumerate}

\section{Experiment Setup}
\label{sec:setting}

The dataset provided for Track 1 of the DiCOVA challenge is a subset of the Project Coswara database\cite{sharma2020coswara} and contains a total of approximately 1.36 hours of cough sound recordings from 75 COVID-19 positive subjects and 965 COVID-19 negative subjects. In addition, the organizers of this challenge adopt a blind test set containing 233 samples, so the results obtained in this paper on the test set are given by the organizers of the challenge.

For evaluating the proposed model performance, the metric have been used is the AUC. To adhere to the rules of the DiCOVA competition, the evaluation parameters have been calculated based on the mean confusion charts from the 5-fold cross-validation.

The model was trained and evaluated using the stratified k-fold cross-validation technique \cite{zeng2000distribution} with 5 folds, similar to the number of folds provided in the challenge. 
The Adam optimizer \cite{kingma2014adam} was used with an initial learning rate of 0.0001 while training on the examples which were further divided into mini-batches of size 32, to implement mini-batch gradient descent with respect to the binary cross-entropy loss function. The training took place over 200 epochs per fold. In the randomness experiment, we only change the random seeds at the beginning of the pipeline. For better training, the order of the training samples is shuffled randomly after every epoch. All the model training and evaluation tasks in this work have been executed in a system with Google Colab, 16 GB RAM, and NVIDIA Tesla V100 GPU.

\section{Results}
\label{sec:results}

\begin{table}[!t]
    \renewcommand{\arraystretch}{1.3}
    \caption{Model AUC scores on the blind test set}
    \label{table:auc}
    \centering
    \begin{tabular}{l||c}
    \hline
    \bfseries Model Architecture & \bfseries Test AUC \\
    \hline\hline
    Random Forest (baseline) & 70.69 \\
    MLP         & 56.57                                 \\
    LSTM        & 70.67                                 \\
    CNN    & 72.33                                 \\
    \textbf{CNN with Data Augmentation}       & \textbf{87.07}  \\
    \hline
    \end{tabular}
\end{table}

As shown in Table \ref{table:auc}, among the four models, our proposal architecture gave the best performance on the cough sounds. Base on the class imbalance problem, we proposed our ConvNet combine Data Augmentation. This helped to improve the AUC score percentage from 72.33\% to 87.07\%, outperformed the baseline model by 23\%.

\begin{figure}[!t]
	\centerline{\includegraphics[width=\linewidth]{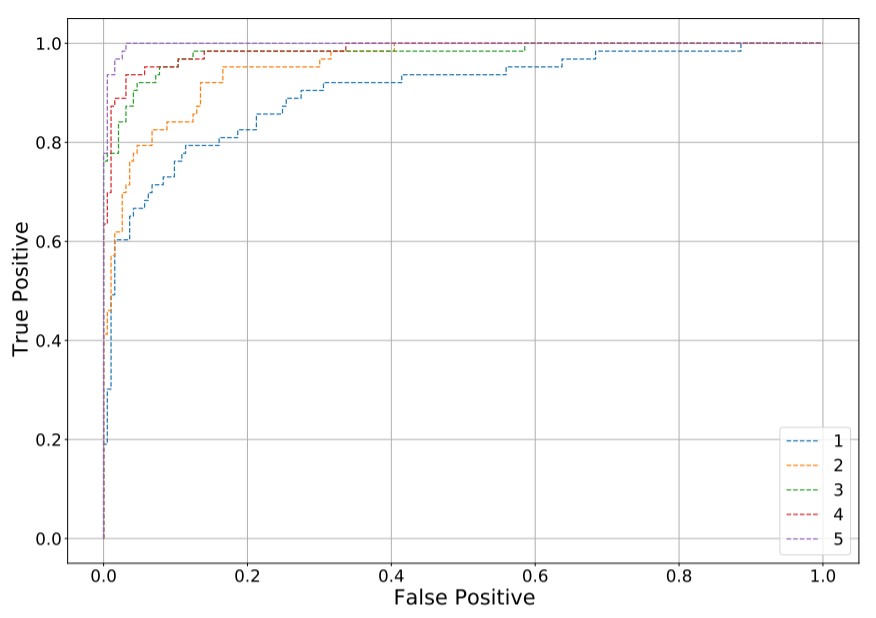}}
    \caption{ROC curves depicting model performance on each fold}
    \label{fig:roc}
\end{figure}

Fig.~\ref{fig:roc} depicts the receiver operating characteristic (ROC) \cite{bradley1997use} curves obtained from the model evaluation on each fold. The validation accuracy averaged over all five folds resulted in 94.61\% with a standard deviation of 2.62\%. In a similar manner, a mean ROC-AUC score percentage of 97.36 was achieved over the folds. 

\begin{figure}[!t]
	\centerline{\includegraphics[width=\linewidth]{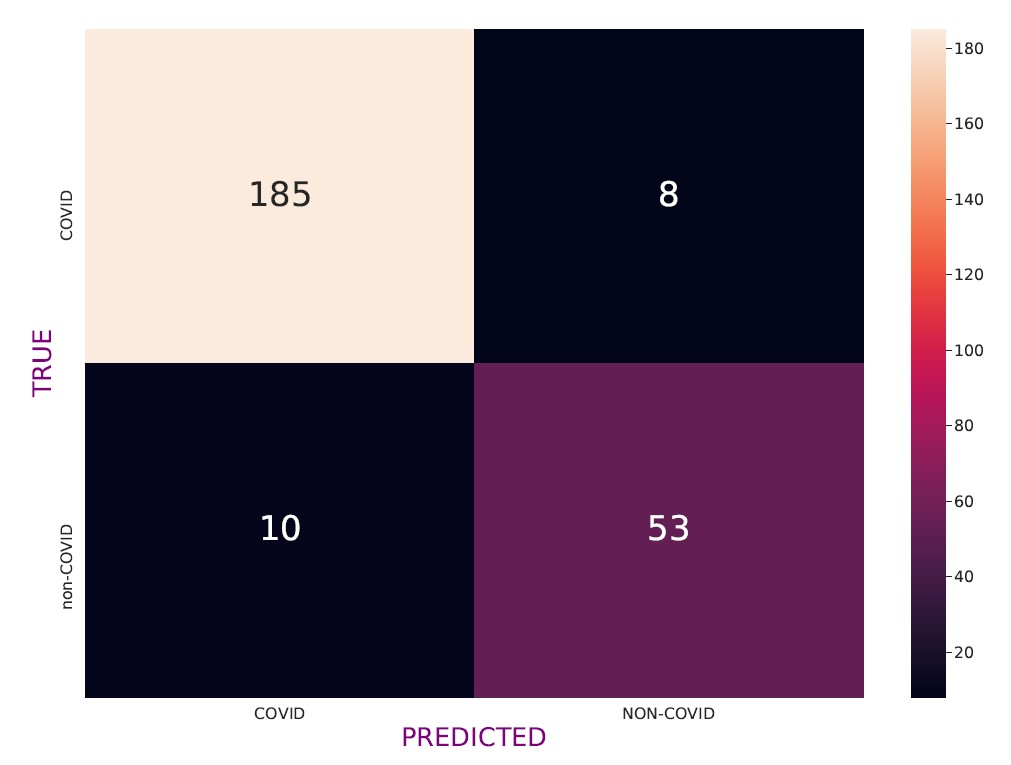}}
    \caption{Averaged model decisions computed at 80\% sensitivity}
    \label{fig:cm}
\end{figure}

Confusion matrices for each fold were evaluated corresponding to the decision threshold giving 80\% sensitivity. Averaging these confusion matrices over the folds for each model, the approximate confusion matrix that was obtained as shown in Fig.~\ref{fig:cm}.

Lastly, the proposed ConvNet model with Data Augmentation achieved a Test AUC score of 87.07 on the blind test set, hence securing the top rank of the DiCOVA 2021 challenge leaderboard \footnote{Challenge Leaderboard: \url{https://competitions.codalab.org/competitions/29640\#results}}. This can be claimed as a truly unbiased evaluation of the proposed model.

\section{Conclusion and Future Work}
\label{sec:conclusion}
With just the limited amount of data that was provided, and some data augmentation techniques, the proposed ConvNet model gave a commendable performance on previously unseen data. It achieved a Test AUC score of 87.07 on the blind test set which outperformed the baseline model by 23\%. Our model is comparatively lightweight since no transfer learning techniques involving a large number of pre-trained were used.

Possible further developments in this area may be the mixture of different sound representations \cite{acoustics2021} such as Chroma Feature, Spectral Contrast, Tonnetz, etc. Additional features from breathing sound may be extracted and interlaced. Such features may be able to formulate distinguishing patterns in the coughing characteristics among patients infected with respiratory diseases like asthma. Using more features may tremendously improve the model performance provided it is trained on a larger dataset with a more balanced class distribution to avoid variance. The proposed model architecture may be trained on a database that is constantly being updated and integrated with web-based applications to boost the process of diagnosing COVID-19, even amongst asymptomatic individuals.

\section*{Acknowledgment}

We would like to thank the organizers of the DiCOVA 2021 Challenge for conducting this impactful competition and providing the dataset. We are also grateful to NIT Silchar for supporting us to carry out the work.

\end{document}